\documentclass[aps,prl,twocolumn,groupedaddress,showpacs,floatfix,letterpaper]{revtex4}

\usepackage{amsfonts}
\usepackage{amssymb}
\usepackage{amsbsy}
\usepackage{amsmath}
\usepackage{amstext}
\usepackage[mathscr]{eucal}

\usepackage[dvips]{graphicx,color}

\usepackage{verbatim}

\newcommand{\lb}[1]{\left #1}
\newcommand{\rb}[1]{\right #1}

\newcommand{\dd}{\mathrm{d}}

\newcommand{\rhom}{\rho_{-}}
\newcommand{\rhop}{\rho_{+}}
\newcommand{\phim}{\phi_{-}}
\newcommand{\phip}{\phi_{+}}

\newcommand{\srhob}{\sigma(\bar{\rho}_+,\bar{\rho}_-)}

\begin{document}

\title{Exact vortex nucleation and cooperative vortex tunneling in dilute Bose-Einstein Condensates}

\author{M.~I.~Parke, N.~K.~Wilkin, J.~M.~F.~Gunn and A.~Bourne}

\affiliation{School of Physics and Astronomy, University of
Birmingham, Edgbaston, Birmingham. B15 2TT. U.~K.}

\begin{abstract}
With the imminent advent of mesoscopic rotating BECs in the lowest
Landau level (LLL) regime, we explore LLL vortex nucleation. An
exact many-body analysis is presented in a weakly elliptical trap
for up to $400$ particles. Striking non-mean field  features are
exposed at filling factors $\gg1 $. {\em E.~g.~} near the critical
rotation frequency pairs of  energy levels approach each other with
exponential accuracy. A physical interpretation is provided by
requantising a mean field (MF) theory, where $1/N$ plays the role of
Planck's constant, revealing two vortices {\em cooperatively}
tunneling between classically degenerate energy minima. The tunnel
splitting variation is described in terms of frequency, particle
number and ellipticity.

\end{abstract}
 \maketitle

The physics of vortices in slowly rotating degenerate gases
\cite{fetter_review} has reached the level of maturity where it is
now used as a tool to study other phenomena, such as polarised fermi
gases\cite{Ket06}. However achieving rapid rotation - to explore
thoroughly the MF quantum Hall (QH)
regime\cite{butts-signatures,ho-many-vortices,nigread,cornell} in
the lowest Landau level (LLL)  \cite{WGS} and to reach correlated QH
states\cite{WGS,WG,CWG} - remains a challenge.

A promising approach to accessing the QH  regime is to have very
dilute BECs, perhaps constructed by slicing up a condensate with an
optical lattice\cite{Dalib}. In this Letter we show that even well
away from the correlated regime there are pronounced quantum effects
which become increasingly striking as the particle number decreases.
We will show that the exact many-body ground states may be
interpreted as exhibiting vortex tunneling leading to superpositions
of mean-field states with vortices residing at different locations.
This mesoscopic limit is consistent with the thrust of experimental
effort in the near future. (In terms of $\nu=N/N_v$, where $N$ is
the number of particles, and $N_v$ the number of vortices, $\nu=1/2$
corresponds to the Laughlin state, and we will study $10 \lesssim
\nu \lesssim 400$.)

Vortex nucleation\cite{ghosh} has been studied in the Thomas-Fermi
regime, both
experimentally\cite{madison-vortex-formation,chevy,hodby-vortex-nucleation}
and theoretically\cite{sinha1,tsubota}. The conclusion is that under
adiabatic ramping of the rotation
frequency\cite{sinha1,chevy,hodby-vortex-nucleation} the process is
determined by an hydrodynamic instability. Under those conditions,
the thermodynamic instability to vortex entry is apparently
unobservable, occurring at lower rotation frequencies.

It is known\cite{RAS} that in a BEC in the LLL in an
axisymmetric(AS) trap that there is a first-order thermodynamic
instability to vortex entry (with no hydrodynamic instability
needed). In this Letter, we will show that the situation is very
different in a non-AS trap. The equilibrium of vortices in a non-AS
trap has already been analysed at a MF
level\cite{linn-elliptical,svidinsky2000pra,oktel-anisotropic}
within the LLL\cite{fetter2007} and at the Bogoliubov
level\cite{sinha2}).

Our starting point is the standard model Hamiltonian, ${\mathscr
H}$, for a cold gas of $N$ particles residing in a plane:
\begin{equation}{\mathscr H}_1\!\!= \!-{\textstyle \frac{1}{2}}\sum_{n=1}^{N}\!\nabla^2_n
+ {\textstyle \frac{1}{2}}\!\!\sum_{n=1}^{N}\! r^2_n + {\textstyle
\frac{1}{2}}\eta \!\!\!\! \sum_{n\ne n'=1}^{N}\!\!\!\!\delta({\bf
r}_n-{\bf r}_{n'}) - \Omega \!\sum_{n=1}^{N}\!L^z_n\nonumber
\end{equation}
Units of length, $a_\perp$, and energy, $\hbar \omega_\perp$, are
those provided by the harmonic trap; angular momenta, $L^z_n$ are
scaled by $\hbar$. There are two remaining dimensionless parameters.
Firstly, $\Omega$, is the angular velocity of the rotating frame
divided by the natural frequency of the harmonic trap. Secondly the
coupling constant, $\eta = 4\pi a/a_\perp$. We assume the particles
reside in the LLL of this AS trap\cite{WGS}. The LLL single-particle
basis utilises a complex description of the particle positions in
the plane, $z=x+ {\rm i}y$, and is defined by the set
$\{(z^m/\sqrt{\pi m!})\thinspace {\rm e}^{-|z|^2/2}\}$, where
$m=0,1,\cdots$.

In the AS case the many-body eigenfunctions, $\psi_L(\{z_k\})$,
(labeled by the total angular momentum, $L$) are known analytically
for $L \le N$ \cite{RAS,WGS,Papenbrock}. They are
$$ \psi_L(\{z_k\}) \propto \sum_{1 \le i_1 \cdots i_L \le N}
(z_{i_1}-z_{\rm c})(z_{i_2}-z_{\rm c}) \cdots(z_{i_L}-z_{\rm c})$$
where $z_{\rm c}=1/N\sum_{j=1}^Nz_j$ is the centre of mass
co-ordinate. The energies, $E_L$, are also known: for $0 \le L \le N
(L \ne 1)$,($L\!=\!1$ is special as it corresponds to centre of mass
motion and the energy is $E_1=1-\Omega$),
\begin{eqnarray}E_L&=&N+L+{\textstyle  \frac{1}{2}}\eta
N(N-{\textstyle  \frac{1}{2}}L-1)-\Omega
L.\label{eq:eng}\\
&=&N+{\textstyle\frac{1}{2}} \eta N(N-1) + (\Omega_{\rm c}^{(0)}
-\Omega)L\label{eq:eng1}.\end{eqnarray} At the critical frequency,
$\Omega_{\rm c}^{(0)}\!=1-\frac{\eta N}{4}$, all the eigenstates for
$2\leq L\leq N$, and $L\!=\!0)$, are degenerate. The entry is
abrupt, but degenerate.

In this Letter we determine how that degeneracy is lifted in a
weakly elliptical trap with striking physical consequences. The
elliptical perturbation has the form: $\frac{1}{4}\epsilon
\mathscr{H}_2=\frac{1}{4}\epsilon\sum_{n=1}^N \lb( z_n^2 + z_n^{*2}
\rb)$. We will assume that $\epsilon \ll \eta$, to allow a
description in terms of the AS states.

\begin{figure}
\includegraphics[width=0.4\textwidth]{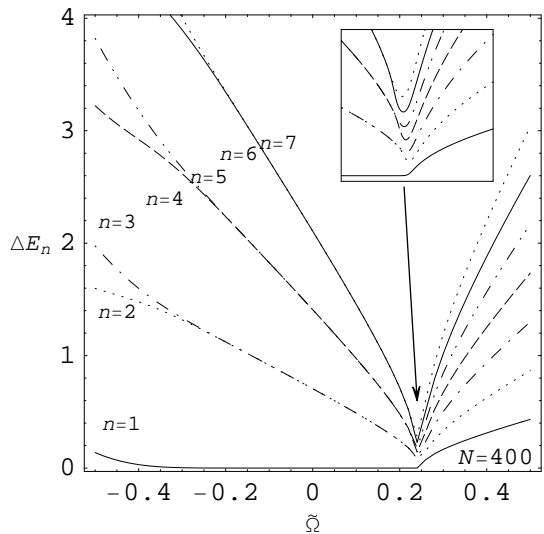}
\caption{\label{fig:energy-gaps} The low-lying energy gaps, $\Delta
E_n$, measured from the ground state for $N=400$ when the trap is
rotated in the vortex formation region near $\tilde{\Omega}\sim 0$.
Inset: detail of the low-lying energy gaps near $\Omega_{c}$, when
the degeneracies lift.}
\end{figure}

To expose clearly the behaviour in the vicinity of vortex
nucleation, it is convenient to use equation (\ref{eq:eng1}). Then,
changing the zero of energy to absorb the term independent of $L$,
we find the complete rescaled Hamiltonian (choosing $\epsilon
>0$ as the unit of energy)
$${\mathscr H}_{\rm tot} = -\tilde{\Omega} {\hat L} + {\textstyle \frac{1}{4}}{\mathscr H}_2$$
where $\tilde{\Omega}= (\Omega -\Omega_{\rm c}^{(0)})/\epsilon$.
This rescaling stretches out the nucleation region where our
approximation is valid ($\epsilon \ll \eta$) which is convenient
numerically. When the energy levels are calculated by exact
diagonalisation (ED) for $N=400$, we find the results portrayed in
Fig.\ref{fig:energy-gaps} (the energies are measure relative to the
ground-state energy and we scale $\eta = \eta_0/N$ to provide a
sensible thermodynamic limit). We restrict ourselves to even $N$, as
for odd $N$ there is a trivial residual first order transition (due
to the necessary change of parity of the ground state) whose
magnitude diminishes as $N\to\infty$.

To understand the significance of the results, it is useful to
consider what the equivalent diagram would have shown in the AS
case. There, from equation (\ref{eq:eng}), we should measure the
energy from the ground state $E_{\rm g} = E_{L=0}$ for $\Omega
<\Omega_{\rm c}$, and from $E_{\rm g} = E_{L=N}$ for $\Omega
>\Omega_{\rm c}$, i.e. the appropriate ground states (within the
basis set in the latter case at higher $\Omega$). Then we find:
\[E_L-E_{\rm g}=\left\{\begin{array}
{r@{\quad:\quad}l} (\Omega_{\rm c}-\Omega)L & \Omega_{{\rm
c}}-\Omega>0\\
(\Omega-\Omega_{\rm c})(N-L) & \Omega-\Omega_{\rm c} >0
\end{array} \right.\]
This would also give a ``V''-shape, however there are substantial
differences. Firstly, in the AS case, the apex of the ``V'' is at
$\Omega - \Omega_{\rm c}=0$, whereas the elliptical case is
displaced by roughly $\epsilon/4$. Secondly, in the AS case the
gradients increase linearly with level index. However in
Fig.\ref{fig:energy-gaps} the energy levels form doublets (which are
of opposite parity) as they approach the critical point, and the
doublets themselves do not have simply related gradients from
doublet to doublet. Of course, the expanded scale of the figure (due
to $\Omega$ being scaled by $\epsilon$) emphasises this and at a
sufficient distance from the critical point, the AS gradients must
be obtained.

\begin{figure}
\includegraphics[width=0.4\textwidth]{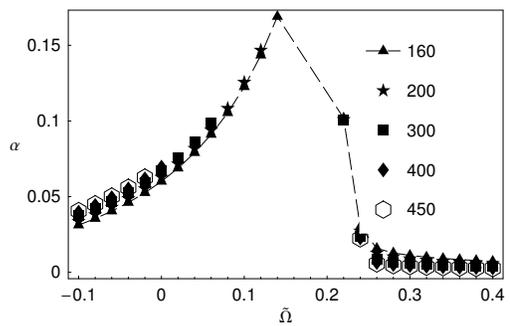}
\caption{\label{fig:alpha1} The tunneling coefficient $\alpha$,
fitted to $N^\beta \,{\rm e}^{- \alpha N}$   from 160-450 particles.
On the left hand side of the peak  $\beta \simeq 0.53$ but the right
hand side does not scale simply. }
\end{figure}

Focussing on the minimum (as a function of $\tilde \Omega$) gap, we
have performed a finite size scaling fit of the gap between the
ground state and first excited state, in terms of the dependence on
the number of particles, $N$, to $\Delta E=N^ \beta e^{-\alpha N}$.
This form is consistent with a tunneling process (which the doublet
structure suggests) occurring prior to the vortex being in the
centre of the BEC. Data for $\Delta E \ge 10^{-14}$ has been kept
for a range of $N$ and a fit found for $\alpha$, as seen in
Fig.\ref{fig:alpha1}.

The most significant feature, is a peak in $\alpha({\tilde
\Omega})$, whose height appears constant with increasing $N$. Having
found the behaviour of $\alpha$, we find that the choice of the
exponent $\beta \approx 0.53$ collapses the data for all $N$ onto
one curve. On the right hand side, there appears to be no single
choice for $\beta$, to collapse the data onto one curve.

The above analysis suggests, via the exponential dependence on $N$
and the pairs of opposite parity eigenstates, that tunneling is
involved in the nature of the states near $\Omega_{\rm c}$. By
analogy with a particle in a double well, the states linked by
tunneling can be exposed by examining the states $\psi_\pm =
1/\sqrt{2} (\psi_0 \pm \psi_1)$, which would correspond to states,
in the double well case, where the particle is localised in one or
the other well. The single particle densities, $\rho_\pm$,
associated with these states, reveal a depression of the density
 on {\em one} side (which is determined
by which of the states is being considered) of the semi-major axis.
This depression in density moves towards the centre of the trap as
$\Omega \to \Omega_{\rm c}$. Further analysis of
Fig.~\ref{fig:energy-gaps} indicates that  gaps between {\em higher
pairs} of opposite parity levels also vanish asymptotically as $N\to
\infty$.

To establish a physical interpretation of these results, we extend
the variational LLL MF Lagrangian\cite{bourne06}, to the elliptical
case. We work with two vortices, as this is the simplest system with
enough parameters to encapsulate the qualitative features of the
exact solution. Following \cite{bourne06} we use hydrodynamic
variables, $\{\rho_m,\phi_m\}$, such that the variational state
$\psi_M(z,z^\star)=\sum_{m=1}^M a_m z^m \rm{e}^{-|z^2|/2}$ and
$a_m=\sqrt{\rho_m/(\pi m!)} \rm{e}^{i \phi_m}$. The {\em two}-vortex
Lagrangian corresponds to maximum angular momentum $M=2$. In units
of $\epsilon$, the two vortex Lagrangian becomes (in the first line
in unscaled form):
\begin{multline}
{\mathscr{L}  =  \hbar\thinspace N\left(\rhom \dot{\phim} +
\rhop \dot{\phip}\right)-\mathscr{H} \label{eq:Lag}}\\
\frac{\mathscr{H}}{\epsilon}  = \lb(\Gamma -{\tilde\Omega} \rb)\lb
(1-\rho_{-}\rb ) +
\sigma(\rho_+,\rho_-)\cos (2\phi_{-}) \\
    + \Gamma
\left (
           {\textstyle\frac{3}{4}}\rho_{-} + {\textstyle\frac{3}{2}}\rho_{+}
        + {\textstyle\frac{3}{32}}\rho_{-}^2 - {\textstyle\frac{21}{8}}\rho_{+}^2 +
        {\textstyle\frac{1}{8}}\rho_{+}\rho_{-}
        + \right .\\ \left .
        2(1-2\rho_{+})\,\sigma(\rho_+,\rho_-)\cos\phi_{+}
        \right ).
\end{multline}
where we have: transformed to variables $\rho_+  =
\frac{1}{2}(\rho_0+\rho_2)$, $\phi_+   = (\phi_0+\phi_2)$, $\rho_- =
(\rho_0 - \rho_2)$ and $\phi_-   = \frac{1}{2}(\phi_0-\phi_2)$;
defined
$\sigma(\rho_{+},\rho_{-})=(1/(2\sqrt{2}))(4\rho_+^2-\rho_-^2)^{1/2}$
and $\Gamma=\eta_0/ 4 \epsilon\pi^2 \sqrt{2 \pi}$;
and used normalisation $\sum_{i=0}^{2} \rho_i=1$, to eliminate
$\rho_1$. (In the vicinity of vortex nucleation $\rho_1 \simeq 1$
and $\rho_0$ and $\rho_2$ are correspondingly small.)  We have
picked the arbitrary phase such that $\phi_1=0$.

At frequencies, $\Omega\lesssim\Omega_{\rm c}$, consideration of the
Lagrangian indicates two degenerate energy minima. By inspection for
the phase variables, $\phi_\pm$, we see that $\phi_+= \pi$ and
$\phi_-= \pm\pi/2$. Minimising the Hamiltonian numerically with
respect to the $\rho_\pm$ gives the equilibrium values
$\bar{\rho}_\pm$ for a given set of interaction and rotation
parameters. In the original variables, the minima have associated
phase variables $\phi_0=0,\phi_2=\pi$, and $\phi_0=\pi,\phi_2=0$.
This implies that both minima correspond to two vortices on the
$x$-axis (one either side of the origin, not necessarily at equal
distance) and are mirror reflections of each other.Beyond MF one
might expect that the vortices tunnel between these configurations
if they are not identical.

Noting that the potential involving $\cos \phi_-$ is the only term
not multiplied by the large parameter, $\Gamma$, we Taylor expand
all the other terms to harmonic level around the equilibria. We also
evaluate $\sigma$, at the equilibrium values of $\rho_\pm$ leading
to a potential $ V_{\phi_{-}} = \epsilon \srhob\cos\phi_{-}$.

We follow the standard procedure {\em e.g.} Ref.~\onlinecite{RingS},
to examine quantum effects beyond MF theory, and re-quantise the
Lagrangian, which now only involves the $-$ variables,  and start
from Eqn. (\ref{eq:Lag}). The {\em classical} conjugate variables to
$\phi_\pm$ are $p_\pm = N\hbar \rho_\pm$. Thus the commutator for
canonically conjugate variables:
$$[{\hat\phi}_{-},{\hat p}_{-}]=[{\hat\phi}_{-},N\hbar{\hat\rho}_{-}]={\rm
i}\hbar \Rightarrow [{\hat\phi}_{-},{\hat\rho}_{-}]={\rm
i}{N^{-1}}.$$ Hence  the usual quantisation procedure implies: $
{\hat \rho}_- \rightarrow -{\rm i}N^{-1}
\partial/\partial\hat{\phi}_-$  This is clearly
unaltered under the displacement of ${\hat \rho}_- = {\bar \rho}_- +
\delta {\hat \rho}_-$. We note that the quantum effects will vanish
in the limit $N\to\infty$, {\em i.e.} the thermodynamic limit,
consistent with the exact results described above.

Making the operator replacements, leads to
\begin{equation}
\lb[-\frac{1}{2M^{*}}\frac{\dd^2}{\dd\phi_-^2} - N^2\left({\cal E} -
\srhob\cos (2\phi_-)\right)\rb]\Psi(\phi_-) = 0
\end{equation}
where we have defined: $1/M^{*} = \partial^2
T/\partial\delta\rho^2|_{\delta\rho_-=0}$ and ${\cal E} =
(E-V_0)/N$, with $V_0$ being the zeroth order terms from the $T$
expansion. This is a Mathieu equation,
\begin{equation}
\lb[\partial_{xx} + b -2q\cos (2 x) \rb] y(x) = 0
\label{eq:two-vortex-mathieu-equation}\end{equation} and inspection
shows that $b \equiv 2N^2 M^*{\cal E} \ge 0$ and $q \equiv N^2
M^*\srhob \ge 0$. In the limiting case of $q=0$, the Mathieu
functions are simply $\cos(\sqrt{b}\phi_-)$ and
$\sin(\sqrt{b}\phi_-)$, so the AS result is recovered.

From Eqn.~(20.2.31)\cite{AS}, the level splitting (between the first
excited and ground states) of
Eqn.(\ref{eq:two-vortex-mathieu-equation}) is $ \Delta b \sim
\sqrt{\frac{2}{\pi}}\;2^5 q^{3/4}{\rm e}^{-4\sqrt{q}} \underset{q
\rightarrow\infty}{\sim} N^{3/2}{\rm e}^{-\alpha_{2v}\thinspace N}
\label{eq:semicllevelsplit} $ implying that $\Delta E \sim
N^{\frac{3}{4}} {\rm e}^{-\alpha_{2v} N}$ where the tunnelling
coefficient $\alpha_{2v} \equiv 4\sqrt{M^{*} \srhob}$.

Finally reinstating the unscaled $\eta$ we can estimate the tunnel
splitting using say $N=10$, $\epsilon=10^{-3}$ and
$\textstyle{\frac{a}{a_\perp}}\simeq 1$ ({\em i.e.} a chip trap). We
then find $E \simeq 0.5 \hbar \omega_0$, indicating the splitting
should be observable under those assumptions.

\begin{figure}
\includegraphics[width=0.4\textwidth]{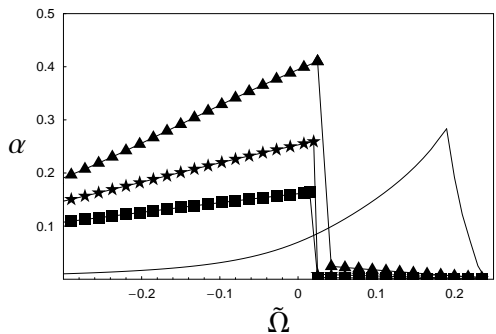}
\caption{\label{fig:alpha} The tunneling parameter from ED (line)
and for two-vortex MF theory with $\Gamma = 50,100,200$
(triangle,star, square). Our approximations become more valid for
increasing $\Gamma$. The MF frequency has been scaled to enable
qualitative comparison of the peaks.}
\end{figure}

These results reproduce the exponential suppression of the tunneling
splitting of the eigenvalues found in the ED, indicating we have
correctly identified the tunneling entities. Qualitatively the
behavior of the tunneling coefficient, $\alpha$, with rotation
frequency (shown in Fig.~\ref{fig:alpha}) follows that of the exact
result, having a maximum in $\alpha (\Omega)$. We have scaled the MF
frequencies, $\tilde{\Omega}$, by $1/4$ to compare the peaks in
$\alpha$. This reflects $\tilde{\Omega}_{\rm c}^{\rm meanf}\simeq
\epsilon$ as against $\tilde{\Omega}_{\rm c}^{\rm exact} =
\epsilon/4$. In addition the exponent of the pre-exponential factor,
$N^\beta$, has $\beta^{\rm meanf}=1/2$ (scaled variables) almost
equal to the ED result, $\beta^{\rm exact} \sim 0.53$, for ${\tilde
\Omega}$ to the left of the peak. We found that adding additional
(up to eight) vortices improves the agreement between the
approximate calculation and the exact one, at the cost of
diminishing the clarity of interpretation (the motivation for the
requantised calculation). Finite temperature will blur the peak, but
leave it visible under typical experimental conditions.

\begin{figure}[h]
\includegraphics[width=0.4\textwidth]{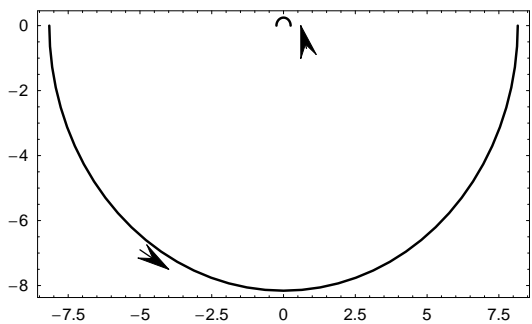}\caption{\label{fig:traj}
Semicircular trajectories of the vortex positions,
ensuring the separation between
vortices is maintained.}
\end{figure}

Working with two vortices the tunneling path may be interpreted in
terms of vortex positions. They are:
$$\xi_{\pm}=-{\rm e}^{{\rm i} \phi_2}\;\sqrt{\frac{\rho_1}{2
\rho_2}} \left ( 1 \pm \left (1+\frac{2 \sqrt{2 \rho_0
\rho_2}}{\rho_1}\right )^{1/2}\right ) $$ Varying $\phi_2$ along
$[0,\pi]$ results in both $\xi_{\pm}$ having semi-circular
trajectories, as shown in Fig.~\ref{fig:traj}. Note, that there is
always a second anticlockwise trajectory corresponding to $\phi_2$
from $[\pi,2 \pi]$, and this will result in the major semi-circle
being above axis.
One might have expected that the tunneling matrix element for a
vortex would increase as the initial and final positions of the
vortex approached each other (i.e. the inner circle in Fig. 4). The
more surprising aspect is that the vortex at large distances from
the centre of the trap does not substantially decrease the matrix
element. This is due to the vortex states (labeled by position)
becoming increasingly non-orthogonal as the vortices move away from
the centre of the trap. The non-orthogonality arises because the
Gaussian weight strongly suppresses the region where the states are
most different \cite{bourne2}. Hence tunneling is least effective
when both vortices are at the same, intermediate, distance from the
trap centre. Exploration of experimental observation schemes is
underway, including oscillation of the trap along the semi-minor
axis and noise spectra which will reveal the density-density
correlation function.

In conclusion, we have shown, by exact solution, that mesoscopic
rotating BECs will show pronounced deviations from MF theory at low
angular velocities. This may be interpreted, using a requantised MF
theory, in terms of vortex tunneling in the nucleation process.

We would like to thank N.R.~Cooper and M.W.~Long for helpful
discussions. This research was supported in part by the NSF
PHY05-51164.

\end{document}